# Viva the *h*-index

Leo Waaijers
leowaa@xs4all.nl

In their article 'The inconsistency of the *h*-index'[1] Ludo Waltman and Nees Jan van Neck give three examples to demonstrate the inconsistency of the *h*-index. As will be shown below, a little extension of their examples just illustrate the opposite, a stable feature of the *h*-index. For starting authors it, the *h*-index that is, focusses on the number of articles; for experienced authors its focus shifts towards the citation scores. This feature may be liked or not but does not make the *h*-index an inconsistent and inappropriate indicator, as the authors claim.

**1. Preamble**
For the sake of compactness of the debate I introduce some terms.
A publication of an author will be called **relevant** if it is one of the *h* publications that constitute his *h*-index. As a corolary, the 'remaining publications' in the definition of the *h*-index[2] are irrelevant[3]. Further, the **production p** of an author is defined as the total number of his relevant publications; the **quality q** of an author is defined as the lowest citation score of his relevant publications.
With these definitions in mind only two cases can happen.
1. The production of an author is lower than his quality: $p < q$.
2. The production of an author equals his quality: $p = q$.
The case $p > q$ does not occur as 'production' is limited by definition to the relevant publications.
Figures 1 and 2 illustrate the respective cases. The notable difference between the two figures is that in figure 1 the 45 degree line intersects the citation-publication curve in a vertical 'production' line of the curve. In figure 2 the 45 degree line intersects a horizontal 'quality' line of this curve.

I now state the following *h*-index 'law'[4]:

$$\text{H-index} = \min(p,q).$$

So, if production p grows quality q becomes the delimiter for the *h*-index. Stated somewhat popularly:

In the *h*-index quality ultimately drives out production.

---

[1] Ludo Waltman, Nees Jan van Eck, The inconsistency of the *h*-index. ArXiv. 19 August 2011. http://arxiv.org/abs/1108.3901
[2] A scientist has an *h*-index of *h* if *h* of their publications each have at least *h* citations and their *remaining publications* each have fewer than *h*+1 citations.
[3] Mind: relevance in this definition is a function of time. Irrelevant publications may become relevant if they are cited more over time; relevant publications may become irrelevant as they are ousted by more cited ones.
[4] Actually, this is not a law but a tautology. Hopefully, it is a helpful one.

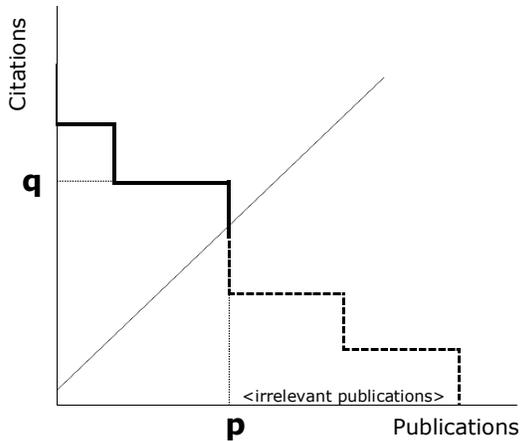
Figure 1: **p**=production, **q**=quality.
Illustrating the case p<q: ***h*-index = p**

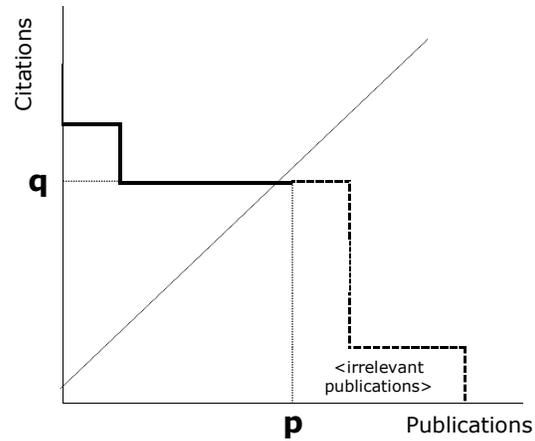
Figure 2: **p**=production, **q**=quality.
Illustrating the case p=q: ***h*-index = q**

## 2. Discussion of the examples
In the discussion of the examples given by Waltman and Van Neck in their article the irrelevant publications will be neglected as they merely have an ornamental function.

**Example 1**
In this example author X has a production of nine with a quality of twelve in five years time. In the same period author Y has a production of seven with a quality of fifteen. So, their *h*-indices are respectively nine and seven. In the next five years they repeat this performance. As a consequence, their *h*-indices go to twelve and fourteen respectively.
Now, suppose they continue this behaviour. So after fifteen years X has written twenty-seven publications with twelve citations each and Y has twenty-one publications with fifteen citations each (under the condition that publications of earlier periods have not gained additional citations in the meantime). In terms of production and quality for X this means p=q=12 and for Y, p=q=15. Hence their *h*-indices are twelve and fifteen respectively. Further continuation of the same achievements does not result in a change of their *h*-indices.
The *h*-indices for X are consecutively 9, 12, 14, 14. For Y they are 7, 14, 15, 15. Apparently, the indices follow the above stated law: in the beginning, when the production is low (i.e. lower than the quality), the *h*-index is determined by the production. When the production grows (and the quality does not) the *h*-index is determined by the quality. That is why X wins in the beginning (higher production) later to be overtaken by Y (higher quality). This is in full accordance with the *h*-index law.

**Example 2**
Here also it is instructive to extrapolate the behaviour of the two scientists X and Y. Originally X has a production of five and a quality of five; Y has a production of four and a quality of six. The *h*-indices of X and Y then equal, respectively, five and four. Next they jointly write two articles which receive eight citations each. The *h*-index of X remains five, the *h*-index of Y goes up to six. If they repeat this step and co-author again two articles that gain eight citations each the *h*-indices increase to six for both authors. Another such step brings their indices to eight. Thereafter the same steps do have no effect no their *h*-indices any more.
Here, the *h*-index history of X is 5, 5, 6, 8 and of Y it is 4, 6, 6, 8. Again, X wins in the beginning based on his higher production. Later Y wins temporarily because of his better quality. Ultimately they end up equally as a consequence of their growing number of joint articles resulting in a 'common' *h*-index. Which is determined by quality. Again, this is in full harmony with the *h*-index law.

**Example 3**
This example demonstrates that the *h*-index of the 'sum' of two authors is not necessarily equal to the sum of their *h*-indices. In the example authors $X_1$ and $X_2$ both have seven publications each with nine citations, resulting in an *h*-index of seven for both authors. When we combine the two authors in a group, the group has fourteen publications that have been cited nine times each. Now, the *h*-index for the group is nine, illustrating again that quality drives out production. A similar configuration holds for the other two authors $Y_1$ and $Y_2$ in this example, with the same conclusion.
If, instead of putting two authors $X_1$ and $X_2$ on the stage this example would have introduced a single author X duplicating his performance and the same would have been done for $Y_1$ and $Y_2$, this example would have matched example 1.

**Group index**
As for *h*-indices of a group the only thing that can be said is that the *h*-index of a group is never lower than the highest *h*-index of their members and never higher than the sum of the *h*-indices of their members. In formula: if the group has n members with respective *h*-indices $h_i$ (i=1,2,..,n) and the group has *h*-index $h_G$ then
$$\max(h_i) \leq h_G \leq \Sigma\, h_i \;(i=1,2,...,n)$$

So, the *h*-index of a group may vary widely. Conversely, widely divergent groups may have the same group index. For example, if a group of ten members has a group-index of ten, then one member may have an *h*-index of ten and the nine others an index of zero, or all members may have an *h*-index of one (each with production one and quality ten), or all members may have an *h*-index of ten. This raises the question of the significance of such a group index. May be the average *h*-index of the group is a more meaningful indicator for assessing a group (or a journal). The examples in the penultimate sentence would then result in respective averages of 0,1; 1 and 10, well reflecting the different composition of the groups.

**3. Conclusion**
The three examples all fulfil the *h*-index law: ultimately quality drives out production. This is an inherent characteristic of the *h*-index; it is in it's DNA. Consequently, the *h*-index praises junior authors for their number of publications but senior authors cannot further increase their *h*-index by more and more publications of approximately the same quality. One might object to this feature and suggest alternative indicators for assessing an author's performance as Waltman and Van Neck do. But inconsistency of the *h*-index is not an argument in this debate. At least, the given examples do not carry that.

Personnally, I like the *h*-index law. It works as in the appraising of walking: with a toddler single steps are applauded for, with an adult it is the tango that makes the difference. Viva the *h*-index.